\documentclass[preprint,showkeys,prc,nofootinbib]{revtex4}

\usepackage{graphicx}
\usepackage{dcolumn}
\usepackage{bm}

\usepackage{natbib}

\newcommand{\kelv}{\ensuremath{\,\mathrm K}}
\newcommand{\teff}{\ensuremath{T_{\rm eff}}}
\newcommand{\abb}[1]{Fig.\,\ref{#1}}
\newcommand{\tab}[1]{Table\,\ref{#1}}
\newcommand{\mdup}{\ensuremath{M_{\rm dup}}}   
\newcommand{\mh}{\ensuremath{M_{\rm H}}}      
\newcommand{\komma}{\mathrm{ \hspace*{0.1cm},}}
\newcommand{\kap}[1]{Sect.\,\ref{#1}}
\newcommand{\jahre}{\ensuremath{\, \mathrm{yr}}}
\newcommand{\lsun}{\ensuremath{\, {\rm L}_\odot}}
\newcommand{\mem}[1]{\ensuremath{\mathrm{ #1}}}
\newcommand{\p}{\ensuremath{\mem{p}}}
\newcommand{\ofu}{\ensuremath{^{15}\mem{O}}}
\newcommand{\czw}{\ensuremath{^{12}\mem{C}}}
\newcommand{\ose}{\ensuremath{^{16}\mem{O}}}
\newcommand{\nvi}{\ensuremath{^{14}\mem{N}}}
\newcommand{\msun}{\ensuremath{\, {\rm M}_\odot}}

\begin{document}

\draft{LA-UR-04-5295}

\title{Nuclear Reaction Rate Uncertainties and Astrophysical Modeling. II. Carbon Yields from Low-mass Giants.}

\author{Falk Herwig} \email{fherwig@lanl.gov}
 \affiliation{Los Alamos National Laboratory, Theoretical Astrophysics
   Group in T-Division, MS B227, Los Alamos, NM 87545}

\author{Sam M.\ Austin}
 \email{austin@nscl.msu.edu}
\affiliation{National Superconducting Cyclotron Laboratory and Joint Institute for Nuclear Astrophysics (JINA), Michigan State University, East Lansing, MI 48824}

\author{John C.\ Lattanzio}
 \email{John.Lattanzio@sci.monash.edu.au}
\affiliation{School of Mathematical Sciences, Monash University, Wellington Road, Clayton, Vic 3800, Australia
}

\date{\today}

\begin{abstract}
  Calculations that demonstrate the influence of three key nuclear
reaction rates on the evolution of Asymptotic Giant Branch stars have
been carried out. We study the case of a star with an initial mass of
2\msun\ and a metallicity of $Z=0.01$, somewhat less than the solar
metallicity.  The dredge-up of nuclear processed material from the
interior of the star, and the yield predictions for carbon, are
sensitive to the rate of the $\nvi(\p,\gamma)\ofu$ and triple-$\alpha$
reactions. These reactions dominate the H- and He-burning shells of
stars in this late evolutionary phase. Published uncertainty estimates
for each of these two rates propagated through stellar evolution
calculations cause uncertainties in carbon enrichment and yield
predictions of about a factor of two.  The other important He-burning
reaction $\czw(\alpha,\gamma)\ose$, although associated with the
largest uncertainty in our study, does not have a significant
influence on the abundance evolution compared to other modelling
uncertainties. This finding remains valid when the entire evolution
from the main-sequence to the tip of the AGB is considered. We discuss
the experimental sources of the rate uncertainties addressed here, and
give some outlook for future work.
\end{abstract}

\pacs{26.20.+fHydrostatic stellar nucleosynthesis, 
  97.30.HkCarbon stars, S stars, and related types (C, S, R, and N),
  97.10.CvStellar structure, interiors, evolution, nucleosynthesis, ages}

\keywords{stellar evolution: AGB stars, nuclear astrophysics: charged particle reactions, individual nuclear reactions: triple-$\alpha$, $\czw(\alpha,\gamma)\ose$, $\nvi(\p,\gamma)\ofu$} 
\maketitle

\section{Introduction}
\label{sec:intro} Modern computer codes for stellar evolution
calculations solve routinely a nuclear reaction network sufficiently
large to account for all relevant nuclear transmutations. Reliable
nuclear reaction rates are a crucial ingredient for accurate modelling
of the evolution of stars, especially for investigations of the
chemical evolution of the stars and of the amount of processed
material that is returned to the interstellar medium. Such stellar
model results are used for integrated models of the galactic chemical
evolution, and for comparison to individual stellar abundance
observations.

In spite of the obvious necessity to check the sensitivity of stellar
chemical evolution predictions to uncertainties in the underlying
reaction rates, little work in this direction has been done recently
\citep{woosley:03}. This is, in particular, true for the evolution of
low- and intermediate mass stars ($0.8<M_\star/\msun<8$) which host
important nuclear production sites and contribute significantly to
galactic chemical evolution. Two main issues make investigations of
the propagation of nuclear reaction rate uncertainties difficult.  One
is the computationally expensive and numerically difficult nature of
the advanced evolutionary phases of low-and intermediate mass
stars. The second is that such studies are only useful if consistent
estimates of the individual rate uncertainties are available.

We have started a program to address this problem. In Paper I
\citep{stoesz:02} we investigated the impact of CNO cycle $(\p,\gamma)$
reaction rate uncertainties on the predicted stellar oxygen isotopic
ratios. These predictions are important for the astrophysical
interpretation of pre-solar meteoritic corundum grains
\citep{huss:94,nittler:97}. In Paper I we used a Monte Carlo approach
integrated into nuclear network post-processing calculations.

In this second study we focus on the evolution of Asymptotic Giant
Branch (AGB) stars and their sensitivity to the rates of three key
nuclear reactions: $\nvi(\p,\gamma)\ofu$, triple-$\alpha$ and
$\czw(\alpha,\gamma)\ose$. We extend our preliminary results of this
project \citep{herwig:04b}, and provide a more in-depth presentation.
 The following sections describe:
\kap{sec:astroback} astrophysical background; \kap{sec:npb} nuclear
physics input, its uncertainties and possible revisions;
\kap{sec:methods} the physical model and methods of the astrophysics
simulation, as well as the main elements of AGB evolution that are
important here; \kap{sec:results} our results as well as additional
calculations with a second, independent code that verify the findings;
\kap{sec:impr} new results for the triple-$\alpha$ rate; and
\kap{sec:discussion} results and discussion.

\section{Astrophysical Context}
\label{sec:astroback}

After the initial H- and He-core burning phases, low- and intermediate
mass stars evolve into double-shell-burning giant stars
\citep{iben:83b,lattanzio:99}. These AGB stars have a unique
mechanical structure, as explained in more detail in
\citet{herwig:04c}.  The electron-degenerate core with mass
$M_\mem{c}\approx 0.6\msun$ consists of carbon and oxygen (the ashes
from previous evolutionary phases) and is about the size of the earth,
while the envelope has on average roughly the density of water and
extends to several hundred times the radius of the sun.  This envelope
is unstable against convection, and is thus well mixed.

Recurrent He-shell flashes, with periods of 5 to 10 $\times
10^{4}\jahre$ are a characteristic of these configurations. A
combination of partial degeneracy, the small geometric scale of the
He-burning shell, and the strong temperature dependence of the
triple-$\alpha$ reaction rate leads to a thermonuclear runaway that
locally generates power of roughly $10^{34}\mem{watts}$ ($10^8\lsun$).
Neither heat conduction nor photon radiation are sufficient to carry
away the energy, and the layer between the He- and the H-shell becomes
convectively unstable.

Part of the energy released in the He-flash will do expansion work,
cooling the layers above the He-shell. The stellar opacities in this
region at the base of the convective envelope then increase, which
forces the envelope deeply into the core (in terms of the Lagrangian
mass coordinate, the enclosed mass). This penetration of the envelope
convection zone into the processed core material below the the
H-burning shell is the third dredge-up.\footnote{It follows two
  dredge-up periods during previous evolutionary phases, which are
  however less important for the overall chemical enrichment of low-
  and intermediate mass stars.} The details of these events are well
documented in the astrophysical literature
\citep{schoenberner:79,iben:81,lattanzio:86,hollowell:88,vassiliadis:93,bloecker:95a}.

The dredge-up event is responsible for the transfer of nuclear
processed material from the high-temperature stellar interior to the
low-temperature stellar surface where it can be observed
spectroscopically. This material is also blown into the interstellar
medium by stellar winds. For these reasons, the strength of the
dredge-up is of great importance to the observed chemical enrichment of
low-mass giants and their role in galactic chemical evolution.

The amount of dredge-up obtained in stellar models of the AGB stars
continues to be a matter of debate. It is well established that the
dredged-up amount depends on the core mass, the stellar metallicity
and opacity, the model of convection, and the treatment of convective
boundaries, as well as on the numerical implementation of several
details in the codes \citep{mowlavi:99,herwig:99a,karakas:02a}.
Earlier studies have already indicated that stronger He-shell flashes
are followed by deeper dredge-up \citep{boothroyd:88}, and that a
decreased energy generation in the H-shell leads to stronger He-shell
flashes \citep{despain:76}. However, there are no investigations of the
sensitivity of dredge-up and the envelope abundance evolution to
nuclear reaction rate uncertainties.

We focus on the three reactions that dominate H-burning and He-burning
respectively. It is well known that the $\nvi(\p,\gamma)\ofu$ reaction
has the smallest rate in the CN cycle; it controls the circulation
rate of the CN catalytic material in the H-burning shell. The
$\czw(\alpha,\gamma)\ose$ reaction is weakly active during the
interpulse phase between the He-shell flashes. The most important
reaction for the He-shell flash is the triple-$\alpha$ reaction.

In double-shell burning around degenerate cores several competing
time-scales are involved. The period of the He-shell flashes depends
mainly on the rate of Helium accretion from the H-burning shell. The
strength of the He-shell flash depends on the geometrical size and on
the partial degeneracy of the He-shell. A more degenerate and thinner
shell results from a longer flash period, which in turn can be caused
by the smaller energy generation rate due to a smaller CNO cycle rate.
In such a case it takes longer to accrete the required amount of He
from the H-burning shell to ignite the flash.  Thus, one can
qualitatively understand that the H-shell burning rate influences the
He-shell flash strength, and thereby the subsequent dredge-up.

The possible influence of the triple-$\alpha$ reaction on the He-shell
flash strength is perhaps more obvious. A larger rate is likely to
cause a larger peak-flash He-burning luminosity, and subsequently a
deeper dredge-up. Our qualitative expectation is then that a reduced
$\nvi(\p,\gamma)\ofu$ rate and an increased triple-$\alpha$ rate will
each increase the amount of carbon produced in the process. The effect
of the $\czw(\alpha,\gamma)\ose$ rate is less obvious. To obtain
quantitative estimates on these processes we have conducted a detailed
numerical study.

\section{Nuclear Physics Input}
\label{sec:npb}
The NACRE collaboration \citep{angulo:99} has recommended reaction
rates for the reactions we consider here. In this section we examine
whether, six years after their publication, the NACRE estimates still
describe the available data with sufficient accuracy for our purposes.
Table \ref{tab:rateselection} lists the temperatures of interest:
$T_\mem{8}=0.5$ ($T=5\times10^7\kelv$) to $T_\mem{8}=0.8$ for the
$\nvi(\p,\gamma)\ofu$ reaction and $T_\mem{8}=1 $--$ 3$ for the
triple-$\alpha$ and the $\czw(\alpha,\gamma)\ose$ reactions.

\subsection{$\nvi(\p,\gamma)\ofu$}
The $^{14}$N($p,\gamma$)$^{15}$O reaction has a complex structure with
transitions, both resonant and non-resonant, to several final states
contributing to the rate. In addition, contributions from the tails of
subthreshold states must be considered.  \citet{schroeder:87} obtained
data over a wide energy range from 0.2 to 3.6 MeV; their R-matrix fit
to this data yielded a large S-factor for the transitions to the
ground state. Their total S-factor, $S(0)= 3.2 \pm 0.54$ keV\,b, was
the principal basis for the NACRE reaction rate $3.2 \pm 0.8
\mem{keV\,b}$ \citep{angulo:99}.  However, the fit to the ground state
cross section required an unusually large value for the gamma width of
a subthreshold $\mem{3/2^+}$ state at $E_x =6.793$ MeV in $^{15}$O,
about 7 times that of the isospin-analog transition in $^{15}$N.  Such
large differences are seldom, if ever, seen, at least for light nuclei
\citep{adelburger:98,brown:04}. Motivated by this fact, direct
measurements of the 6.793 state's lifetime were made
\citep{bertone:01,yamada:04} and yielded much smaller gamma widths.  A
reanalysis \citep{angulo:01} of the Schroeder data resulted in a much
smaller ground state transition and $S(0)= 1.77 \pm 0.2$ keV\,b.
Later, determinations of asymptotic normalization coefficients (ANC)
for the relevant transitions using nuclear transfer reactions,
combined with some of the Schroeder results \citep{akram:03} led to
$S(0)= 1.7 \pm 0.41$ keV\,b.

Recently, additional $^{14}$N($p,\gamma$)$^{15}$O data and corrected
data from \citet{schroeder:87} were analyzed to yield $S(0)= 1.70 \pm
0.22$ keV\,b \citep{formicola:03}.  An independent measurement at TUNL
\citep{runkle:04} yielded $S(0)= 1.68 \pm 0.18$ keV\,b.  Measurements
of the analyzing power at 270 keV \citep{nelson:03} indicate that M1
contributions to the cross section should be considered; to our
knowledge these contributions have not been included in detailed fits
to the data.  There are also preliminary data down to 70 keV
\citep{costantini:04} for the main transition, that to the 6.79 MeV
state.

All the recent investigations show that the ground state transition is
small and that the resulting total S factor is smaller by about a
factor of two than the NACRE result. In the near future, the
reliability of the rate is likely to improve as more complete analyses
including all the recent data are carried out.  However, for now we
have chosen to use an unweighted average of the four recent results
and a conservative error reflecting the relatively long extrapolations
to the astrophysical range, the neglect of the M1 amplitudes, and the
imperfect fits to the data. We obtain $S(0) = 1.70 \pm 0.25$ keV\, b.
For easy employment in the stellar evolution code, this value can be
approximated as a fraction $f = 0.64\pm0.1$ of NACRE's analytical fit
\citep{angulo:99} to the reaction rate in the relevant temperature
range.

\subsection{Triple-$\alpha$}
The first step of the triple-$\alpha$ process is the fusion of two
$\alpha $ particles to form a equilibrium concentration of $^8$Be. The
subsequent capture of an alpha particle produces an equilibrium
concentration of $^{12}$C in its 7.65 MeV $\mem{O^+}$ state.
Occasionally $^{12}$C is formed by a leak, via a gamma cascade or pair
emission, to the ground state of $^{12}$C.  For the temperatures
involved in the present calculations both of these steps are resonant
and the reaction rate is given by:

\begin{equation}\label{3alpha}
r_{3\alpha}\propto \Gamma_{rad}\exp(-Q/kT).
\end{equation}

The value of $Q$ for the 7.65 MeV state is known to within $\pm 0.2$
keV \citep{nolen76} and contributes an uncertainty in the rate of only
$\pm 1.2\%$ for $T_8=2$.  Essentially all the uncertainty in the rate
is due to the uncertainty in $\Gamma_{rad}$, the radiative width of
the 7.65 MeV state; $\Gamma_{rad}$ is known with a precision of $\pm
12\%$. It is essentially these established values of
$\Gamma_\mem{rad}$ and $Q$ that are incorporated into the NACRE rates
for the temperatures considered here, and hence, for our purpose the
NACRE rates are adequate. We shall see, however, that they are not
sufficiently accurate.

Following the completion of the calculations described herein,
\citet{fynbo:03}, \citep{fynbo:05} determined the level structure of
\czw\ and concluded from their results that the 7.65 MeV state alone
adequately describes the reaction rate for temperatures $T = 0.1 - 100
\times 10^{8}\kelv$. They also conclude that at $T = 2\times
10^8\kelv$, the midpoint of the temperature range covered here, the
triple-alpha rate is smaller than the NACRE results by about $10\%$.
This is comparable to the quoted error we used, and should be kept in
mind when examining the details of the present results. In fact, our
calculations can be used to determine the effect of the $10\%$ change
on C-yields.

\subsection{$\czw(\alpha,\gamma)\ose$}
This reaction has been the subject of many experiments and analyses
over a period of forty years, but it is still not accurately known. It
is not possible here to review this subject in detail, it is simply
too complex.  One can find a comparison of the various rates in Fig. 1
of \citet{eleid04}, and the references cited there can be consulted
for more detail. A extensive discussion of this rate can be found in
\citet{buchmann04} and the results of extensive recent measurements in
\citet{hammer04}. The NACRE rate is probably somewhat too large in the
region of present interest, but the quoted uncertainties are
sufficiently large to represent the probable range of acceptable
values.  This is a minor issue for the present calculations since, as
we shall see, carbon production in low and intermediate mass AGB stars
is very weakly dependent on the $\czw(\alpha,\gamma)\ose$ rate.

\subsection{Comments}
We conclude that the NACRE recommendations describe current
triple-$\alpha$ and $\czw(\alpha,\gamma)\ose$ data sufficiently well,
at least for the temperature range relevant here. Recent data,
however, make it clear that the NACRE estimate for the
$\nvi(\p,\gamma)\ofu$ rate is high by roughly a factor of two.

\section{Methods and physics input}
\label{sec:methods}

\begin{table*}
\caption{\label{tab:rateselection} Relevant temperature range, NACRE nuclear reaction rates and their uncertainties \citep{angulo:99}, and adopted factors to fitting formula rates for our calculations.}
\begin{ruledtabular}
\begin{tabular}{lllllllll}
reaction &$T_8$&$<\sigma v>_\mem{low}$&$<\sigma v>$\footnote{Recommended reaction rate.}&$<\sigma v>_\mem{high}$&exp\footnote{Power of 10 multiplying reaction rates in columns 3,4, and 5}& $f_{fit}$\footnote{Ratio between tabulated value and fit formula.}
&$f_{up}$\footnote{$<\sigma v>_\mem{high}/(<\sigma v> f_{fit})$.}
&$f_{low}$\footnote{$<\sigma v>_\mem{low}/(<\sigma v> f_{fit})$}\\ \hline
$\nvi(\p,\gamma)$       &$  0.5    $&$2.67    $&$3.68    $&$4.69  $&$-10  $&$0.9701    $&$1.3137    $&$0.7479 $\\
                        &$  0.8    $&$0.76    $&$1.04    $&$1.32  $&$ -7  $&$0.9488    $&$1.3377    $&$0.7702 $\\
\multicolumn{7}{l}{adopted for our calculations: \dotfill}&$1.33$&$0.75$\\ \hline
$3\alpha$               &$  1.0    $&$2.05    $&$2.38    $&$2.70  $&$-24  $&$1.0424    $&$1.0883    $&$0.8263 $\\
                        &$  3.0    $&$3.95    $&$4.57    $&$5.18  $&$-13  $&$1.0068    $&$1.1258    $&$0.8585 $\\
\multicolumn{7}{l}{adopted for our calculations: \dotfill}&$1.13$&$0.82$\\ \hline
$\czw(\alpha,\gamma)$   &$  1.0    $&$1.06    $&$1.81    $&$2.55  $&$-20  $&$0.9762    $&$1.4431    $&$0.5999 $\\
                        &$  3.0    $&$2.88    $&$4.75    $&$6.62  $&$-12  $&$0.9905    $&$1.4070    $&$0.6121 $\\
\multicolumn{7}{l}{adopted for our calculations: \dotfill}&$1.44$&$0.60$\\
\end{tabular}
\end{ruledtabular}
\end{table*}

The one-dimensional stellar evolution codes we employed solve the
well-established full set of non-linear partial differential equations
to account for hydrostatic equilibrium, mass continuity, and energy
transport and generation
\citep{kippenhahn:90,bloecker:95a,herwig:99a}. Most of the
calculations have been done with the code EVOL \citep{herwig:03c}. It
includes updated input physics.\footnote{The opacities, for example,
  are from \citet{iglesias:96} supplemented with low temperature
  opacities by \citet{alexander:94}.}  A small amount of envelope
overshooting, but no overshooting at other AGB convection zones is
introduced following \citet{herwig:97}.\footnote{The efficiency for
  convective envelope (CE) overshooting is $f_\mem{ce}=0.016$, see
  \citet{herwig:99a} for a description of the overshooting scheme
  used.  Mass loss is included by adopting the formalism of
  \citet{bloecker:95a} with a scaling factor $\eta_\mem{B}=0.1$. For
  more details, and definitions see \citet{herwig:03c}.}

In order to generate the initial model for our comparative study we
compute the evolution of a star with a mass of $2\msun$ and a
metallicity of $Z=0.01$ from the pre-main sequence through the H- and
He-core burning phase. After the He-core burning phase the star
gradually climbs up the Asymptotic Giant Branch in the $\log \teff$ -
$\log L$ (Hertzsprung-Russell) diagram. We choose as a starting model
for all subsequent calculations a model at the very end of the He-core
burning phase, and well before the onset of the first He-shell
flashes.

The evolution of the H- and He-burning shell and of the stellar
surface in the Lagrangian mass coordinate for the benchmark sequence
ET2 are shown in \abb{fig:t-M-ET2-long}. From the end of the He-core
burning phase the star spends about 20 million years on the so-called
Early-AGB phase. During this phase the H-burning shell is largely
inactive and most nuclear energy is produced in the He-burning shell.

He-shell flashes occur only during a rather short period of the
post-He-core burning phase.  The underlying reason for their
occurrence is the different burning rate of the two shells which
eventually prohibits quiescent double-shell burning. A close-up of the
actual He-shell flash phase of the AGB is shown in
\abb{fig:t-M-ET2-short}.  Seventeen thermal pulses can be identified
by the vertical lines that connect the H- and the He-shell at almost
equidistant intervals. These vertical lines are the brief He-shell
flash convection zones which last for only $200$ to $300\jahre$.
During the flash the convectively unstable layers are confined to the
region below the H-shell and above the He-shell.  The inset shows a
small spike at the bottom which is the rapidly growing upper boundary
of the He-shell flash convection zone. It stops just short of the mass
coordinate of the H-free core where the H-shell is located, and
quickly retreats afterward. The H- and He-burning shells remain well
separated even during the He-shell flash episodes, and no H from the
envelope can enter the He-burning shell. The inset also shows how the
bottom of the convective envelope later descends into mass layers
previously occupied by the He-shell flash convection zone and
with consequent ``dredging'' of
processed material into the envelope. It is this tiny detail in the
convective evolution of the stellar interior that is responsible for
the enrichment of the envelope and eventually, through mass loss, of
the interstellar medium. The dredge-up events after the thermal pulses
cause a gradual increase of carbon and to a much lesser extent oxygen
(\abb{fig:t-CO}).  The surface abundance for $^{16}$O is nearly
constant with time for near-solar metallicity. This results in an
increase of the C/O ratio, and eventually to the formation of C-stars.
More detailed figures of the evolution of He-shell flashes can be
found, for example, in Fig.~1 and 10 in \citet{herwig:99a}.

For any comparative study of the propagation of nuclear reaction rate
uncertainties in a stellar evolution code a somewhat consistent set of
quantitative estimates on the uncertainties is required. In the
absence of such estimates, one is left with the rather crude approach
of applying common factors to all rates in question. However, such an
approach misses out on critical aspects of the error propagation in a
real stellar evolution environment. As we will show, for example, the
reaction in our sample with the largest relative error has the
smallest impact on the observable prediction. Similar conclusions were
drawn in Paper I \citep{stoesz:02}.

In this study we rely initially on the NACRE compilation
\citep{angulo:99} which contains recommended values and estimates for
lower and upper bounds as a function of temperature. We then consider
the impact of the revised recommendation for the $\nvi(\p,\gamma)\ofu$
reaction. Note that the recommended factors only apply in the given
temperature range.  In \tab{tab:rateselection} and \abb{fig:overview}
we show the relevant information from the NACRE compilation for the
two temperature ranges appropriate for He- and H-burning respectively.
In addition to the tabulated reaction rates, fitting formula for the
recommended values are provided. In the calculations we use these
formulae instead of tables to evaluate the reaction rates at the
required temperatures. We have checked the accuracy of the formulae,
and found that for the $T$ range of interest here (as indicated in
\tab{tab:rateselection}, column 2) the fitting error is rather small
as shown in Column 7 of \tab{tab:rateselection} which shows
$f_\mem{fit} = <\sigma v>_{table}/<\sigma v>_{fitformula}$.

\section{Calculations and Results}
\label{sec:results}

\subsection{Chemical enrichment and dredge-up as a function of nuclear physics input}
\label{sec:calc1}
\begin{table*}
\caption{\label{tab:sequences1} Results for NACRE recommended rates and uncertainties.}
\begin{ruledtabular}
\begin{tabular}{cccccccc}
ID& reaction &factor & $N_\mem{TP}$\footnote{Number of Thermal Pulses that cause dredge-up.}&
$\lambda(m_\mem{c}=0.56\msun)$\footnote{Dredge-up efficiency $\lambda$ at mass coordinate $0.56\msun$, for details see text.}&
$\lambda_\mem{max}$\footnote{Maximum $\lambda$ reached by any  flash in  the entire sequence.}&
$\sum \mdup/10^{-2}\msun$\footnote{Entire mass dredged-up by all dredge-up events.}&
$p_{^{12}\mem{C}}$ \footnote{\czw\ yield from thermal pulses and the accompanying dredge-up in units of $10^{-3}$\msun. For details see text.}  \\ \hline
ET2& all               &$ 1.00$&$ 8$&$ 0.16$&$ 0.29$&$1.2$&$2.19$\\
ET5& $\nvi(\p,\gamma)$ &$ 1.33$&$ 9$&$ 0.15$&$ 0.31$&$1.2$&$1.95$\\
ET8& $\nvi(\p,\gamma)$ &$ 0.75$&$11$&$ 0.29$&$ 0.39$&$2.3$&$4.27$\\
ET6& $3\alpha$ &$ 1.13$&        $10$&$ 0.31$&$ 0.41$&$2.5$&$5.42$\\
ET9& $3\alpha$ &$ 0.82$&        $ 7$&$ 0.12$&$ 0.29$&$1.1$&$1.79$\\
ET7& $\czw(\alpha,\gamma)$&$  1.44$&$ 9$&$ 0.24$&$ 0.34$&$1.6$&$2.71$\\
ET10& $\czw(\alpha,\gamma)$&$ 0.60$&$ 8$&$ 0.21$&$ 0.33$&$1.5$&$3.12$\\
\end{tabular}
\end{ruledtabular}
\end{table*}
\begin{table*}
\caption{\label{tab:sequences2} Results for revised $\nvi(\p,\gamma)\ofu$ range.}
\begin{ruledtabular}
\begin{tabular}{cccccccc}
ID& $\nvi(\p,\gamma)$ &  $3\alpha$ & $N_\mem{TP}$\footnote{See \tab{tab:sequences1} and text for explanations and details.}&$\lambda(m_\mem{c}=0.56\msun)$&$\lambda_\mem{max}$& $\sum \mdup/10^{-2}\msun$&$p_{^{12}\mem{C}}$ \\ \hline
ET12& $ 0.75$&$ 1.13$&$10$&$ 0.33$&$ 0.43$&$2.5$&$5.65$\\
ET13& $ 0.64$&$ 1.00$&$ 9$&$ 0.29$&$ 0.41$&$2.2$&$4.62$\\
ET14& $ 0.64$&$ 1.13$&$11$&$ 0.36$&$ 0.44$&$2.9$&$6.02$\\
ET15& $ 0.54$&$ 1.13$&$11$&$ 0.41$&$ 0.46$&$3.2$&$6.98$\\
ET17& $ 0.54$&$ 0.82$&$ 9$&$ 0.25$&$ 0.39$&$2.3$&$5.29$\\
ET18& $ 0.75$&$ 0.82$&$ 8$&$ 0.16$&$ 0.41$&$1.8$&$3.98$\\
\end{tabular}
\end{ruledtabular}
\end{table*}


Starting from our initial model at the end of core He-burning, we
calculate seven full evolutionary sequences which end when all
envelope mass is lost and the remaining stellar core is about to
become the central star of the planetary nebulae stage. The sequences
differ only in the adopted rates for the three reactions investigated
here (\tab{tab:sequences1}). One benchmark sequence (ET2) has been
computed with the recommended NACRE rates for all three reactions. In
addition six sequences have been calculated in which for each of these
reactions the rate from the fitting formula is multiplied by the
factors given in \tab{tab:rateselection}, which include the small
differences between tabulated and fitting formula values. Thus, for
each reaction two sequences are computed, one adopting the upper and
one adopting the lower bound of the uncertainty range.

In a second set of sequences we have investigated the influence of the
revised (smaller) $\nvi(\p,\gamma)\ofu$ rate.  We have considered also
simultaneous changes of both this rate and the triple-$\alpha$ rate
(\tab{tab:sequences2}). An overview of the reaction rate choices for
the most interesting cases in which the $\nvi(\p,\gamma)\ofu$ and the
triple-$\alpha$ rate have been changed is shown in \abb{fig:overview}.

We summarize these results in \tab{tab:sequences1} and
\tab{tab:sequences2}. The most salient features of these results are
described by the efficiency of dredge-up, $\lambda$, and the
  yield of carbon $p_{^{12}C}$.  The efficiency is given by
  $\lambda=\Delta \mdup / \Delta \mh$ with $\Delta \mh$ the core mass
  growth between two He-shell flashes due to H-shell burning, and
  $\Delta \mdup$ the dredged-up mass following the He-shell flash. For
  each flash $\lambda$ indicates the efficiency of dredge-up.
  $\lambda=1$ means that the same amount of mass by which the core
  grew between two flashes is dredged-up after a flash. The $^{12}$C
yield from thermal pulses and the accompanying dredge-up is given for
$i=^{12}$C by: $p_i = \int_{M_\mem{f}}^{M_\mem{i}}
(X_i(m)-X_\mem{ini})\, \mem{d}m \komma$ where $M_\mem{i}$ and
$M_\mem{f}$ are the initial and the final stellar mass at the
beginning and the end of the AGB phase, $X$ is the mass fraction at
the surface as the star evolves and $X_\mem{ini}$ is the initial mass
fraction. We also tabulate the total dredged-up mass.
 
The chemical evolution of AGB giants depends sensitively on the rates
of the $\nvi(\p,\gamma)\ofu$ and the triple-$\alpha$ reaction.
Calculations with a smaller $\nvi(\p,\gamma)\ofu$ rate (case ET8) show
a larger dredge-up efficiency, a larger dredged-up mass, and a larger
amount of carbon mixed from the processed layers to the envelope and a
larger carbon yield.\footnote{The carbon enrichment is used here as a
  proxy for the envelope enrichment with nuclear processed material
  which would include the s-process elements, for example.}  All these
quantities are about a factor of two higher than for the benchmark
case.

For the triple-$\alpha$ reaction we observe the opposite behavior. The
case ET6 with a larger rate has, on average, He-shell flash peak
luminosities (not shown in the table) that are 20-50$\%$ higher than
the benchmark case and accordingly dredge-up is deeper. The
efficiency, total dredged-up mass, and the \czw\ yield are about a
factor of two higher for sequence ET6 compared to sequence ET2.  It is
also clear that these effects are non-linear in the rates.  The
changes for increases in the $\nvi(\p,\gamma)\ofu$ rate and decreases
in the triple-$\alpha$ rate are small. 

The uncertainty in the $\czw(\alpha,\gamma)\ose$ as given in the NACRE
compilation rate has a much smaller influence on the observables
studied here. We find that in both cases with upper and lower range
values for this rate (cases ET7 and ET10) the \czw\ yield is somewhat
larger than in the benchmark case. Sequence ET10 has a slightly larger
\czw\ yield than ET7 although ET7 has a slightly larger dredge-up mass
and efficiency than ET10 (see discussion in \kap{sec:model-checks}).

In \tab{tab:sequences2} we show the results for the revised
$\nvi(\p,\gamma)\ofu$ rate and for cases in which two rates are
changed simultaneously.  The $^{12}$C yield increases further for the
lower $\nvi(\p,\gamma)\ofu$ rate and the combination with an increased
triple-$\alpha$ rate leads to still larger values. A compact
representation of the set of simulations is given in
\abb{fig:C12yield}. The main result is that due to the revision of the
$\nvi(\p,\gamma)$ rate the predicted \czw\ yields of low-mass stars is
about twice as large as with the old rate. In addition, the relative
error of the combined effect of the $\nvi(\p,\gamma)$ and
triple-$\alpha$ rate has decreased.

The evolution of the \czw\ abundance is useful to further illustrate
the differences. In \abb{fig:C12} the envelope \czw\ abundance
increases in discrete steps for all cases. These steps correspond to
the discrete dredge-up events after sufficiently strong He-shell
flashes. The astrophysical yield is obtained by integrating the
surface abundance over the mass lost. Most notable is the fact that
\czw\ abundances for the shown cases span a range significantly
exceeding a factor 2.  Until the reaction rates are better known this
is an unavoidable uncertainty in the yield predictions.

These results confirm our original qualitative expectations.  A
smaller $\nvi(\p,\gamma)\ofu$ rate leads to a smaller helium
production rate and later ignition of the He-shell flash. This flash
is then more violent and the subsequent dredge-up is more efficient
compared to a case with a larger $\nvi(\p,\gamma)\ofu$ rate. More
efficient dredge-up leads to a larger envelope enrichment, and thus
the envelope \czw\ abundance and stellar yield is larger for the run
with the smaller $\nvi(\p,\gamma)\ofu$ rate.  For the triple-$\alpha$
reaction a larger rate leads to stronger He-shell flashes. In fact the
run with the large rate shows He-burning peak luminosities which are
about 20-50\%\ larger than for the run with the lower triple-$\alpha$
rate. Accordingly the run with the larger rate shows greater
efficiency, deeper dredge-up, and larger \czw\ abundances at the
surface and in the yields.

\subsection{Verification of stellar modelling results}
\label{sec:model-checks}
The main set of models has been computed using the EVOL stellar
evolution code \citep{herwig:99a}. In order to check these results we
have repeated a subset of the numerical experiments with the
independent MSSSP code \citep{lattanzio:86}. Calculations were carried
out for $M=2.1\msun$ and $Z=0.008$ with the standard $\nvi(\p,\gamma)$
rate used in that code (very similar to the NACRE rate) starting from
the main-sequence and a comparison calculation starting, as with the
EVOL set of sequences, after the end of core He-burning with $0.6$
times the standard $\nvi(\p,\gamma)$ rate. This calculations shows
deeper dredge-up and a larger C yield compared to the sequence with
the higher $\nvi(\p,\gamma)$ rate. The effect seen in the MSSSP
calculations is qualitatively and quantitatively consistent with the EVOL results.

In several test calculations we discovered that starting the
comparison runs with different reaction rates at or even after the
onset of the He-shell flashes did not show consistent trends. The
luminosity of the H- and He-shell depend on the core properties (im
particular mass and radius) that are the result of the previous core
burning phases of H and He. For the main set of comparative
calculations we therfore choose an initial model about 20 million
years before the first thermal pulse, and immediately after the end of the
He-core burning.

Comparison calculations have been done with the MSSSP code with
different $\czw(\alpha,\gamma)\ose$ rates. As with the EVOL
calculations reported in \kap{sec:calc1} we were not able to identify
clear correlations between this reaction rate and the dredge-up and
yield properties of the models. The differences we found were below the
$10 - 15\%$ level and more sensitive to numerical parameters (such as
the spatial and temporal resolution) than the
$\czw(\alpha,\gamma)\ose$ rate. (This is actually to be expected,
because this reaction rate is not as important as the triple-$\alpha$
rate during the AGB thermal pulse evolution.)  The quantities like
total dredge-up mass and carbon yield are the result of a discrete
process repeated $ n \approx 8$ to $11$ times.  Thus, these numbers are
subject to a statistical fluctuation $\sim 1/n$.

\subsection{The pre-AGB evolution: H- and He-core burning}
\label{sec:pre-AGB}

The main emphasis of this study is the impact of nuclear reaction
rates on the AGB evolution, in particular the chemical enrichment
through the third dredge-up. However, we did check how the dredge-up
of models with different reaction rates change if run all the way from
the main-sequence, including the effect on H- and He-core burning. One
test has been done with the MSSSP code, starting a sequence with the
reduced $\nvi(\p,\gamma)$ rate from the main-sequence. In that
sequence the dredge-up during the AGB thermal pulse phase is still
significantly larger than with the NACRE recommended rate, but the
increase is somewhat smaller then for the comparison runs started
after the end of He-core burning.

A second test has been
made with the EVOL code running two sequences all the way from the
zero-age main-sequence to the end of the thermal pulse AGB. In
addition to the benchmark case ET2 we reran the combination of case
ET14 ($0.64\times \nvi(p,\gamma)$, $1.13 \times 3\alpha$), and refer to
this run as ET14a. We found that in this case the dredge-up is about
$20\%$ larger than in the ET14 model calculated from the starting
model after the end of He-core burning. This is another example of the
highly non-linear behaviour of the third dredge-up. 

We can analyse the differences caused by the change of nuclear
reaction rate on the H- and He-core burning phases, and find them to
be very small.  The central temperature during the H-core burning of
the $2\msun$ models studies here are $2.1 \times 10^7\kelv$ initially,
increasing sharply to $3.45 \times 10^7\kelv$ at the end of H-core
burning. Case ET14a shows central temperatures throughout core
H-burning that are $1\%$ larger than the benchmark case. This small
increase of temperature is sufficient to increase energy generation
required for hydrostatic equilibrium because of the steep temperature
dependence of the $\nvi(\p,\gamma)$ rate. Run ET14a consumes H in the
center slightly faster, and accordingly the H-core burning phase is
about $1\%$ shorter than in the benchmark case. The mass of the
convective core is practically the same in both cases. None of these
nuclear reaction rate differences during the main-seqeunce evolution
would make an observable difference.

During the He-core burning the slightly larger triple-$\alpha$ rate
leads to a slightly larger C/O ratio on the core ($2\%$). However,
during the EVOL calculations some breathing pulses of the convective
core occur. These breathing pulses are well known during the He-core
burning phase and related to the unstable growth of the convective
core into a layer that is stabilized by an composition gradient. The
treatment of convective boundaries implemented in the EVOL code makes
the occurrence of breathing pulses somewhat dependent on the numerics.
It is not clear from the calculations to which extent the magnitude of
the central C/O ratio depends on this simulation error. However, since
the effect is small we decided not to follow this question any
further. At the end of He-core burning the age difference between the
two runs is $0.3\%$ of the total age. The core mass and size are
practically identical.

The second dredge-up decreases the core mass slightly for stars of
this mass. This effect is weaker in run ET14a, so that the core mass
after the second dredge-up is slightly larger than in the benchmark
run. It is during the early AGB evolution from the end of He-core
burning to the first thermal pulse that sequence ET14a has a slower
growth rate of the H-free core, corresponding to a $3\%$ smaller
H-burning luminosity. This leads to a $2\%$ smaller core mass at the
first thermal pulse for case ET14a compared to the benchmark run. A
smaller core mass should result in less efficient third dredge-up, if
the reaction rates are the same. However, ET14a has a combination of
rates that increase third dredge-up, as shown in the main set of
comparative calculations. In addition, because the stellar luminosity
of run ET14a is somewhat smaller, the mass loss according to the
adopted L-dependent mass loss formula is smaller, and the thermal
pulse AGB phase is longer by $15\%$. In particular the last thermal
pulses have larger He-flash peak luminosity, that as discussed above
leads to more efficient dredge-up.

The third test concerns the influence of the $\czw(\alpha,\gamma)\ose$
rate on the pre-AGB evolution. We calculated an additional sequence
(ET7a) corresponding to sequence ET7 (NACRE rate times 1.44) from the
main-sequence to the thermal pulse AGB phase. As can be expected,
there are no difference (e.g. duration, core size) between run ET7a
and ET2a during the H-core burning phase.  During the core He-burning
phase the central C/O ratio is systematically smaller in run ET7a
compared to the benchmark case. For example after 1/3 of the He-core
burning phase $\mem{C/O} = 3.6$ for the ET7a run and $\mem{C/O} = 5.2$
for the benchmark case. Not surprisingly the ratio of those two values
is 1.44. The ET2a He-core burning duration is $3\%$ shorter than ET7a,
the central temperature and density are almost the same. As in the
discussion of run ET14a above we note the occurence of breathing
pulses during the He-core burning phase, in particular towards the end
of this phase. In runs ET2a and ET7a these breathing pulses are very
similar, with no observable difference. The core mass at the first
thermal pulse is $0.494\msun$ for ET7a, and practically the same
($0.493\msun$) for ET2a.  We continued run ET7a into the thermal pulse
regime until the onset of the third dredge-up. As expected from the
practically idenitical core masses the ET7a and ET2a sequence show
very similar third dredge-up behaviour.

In conclusion we find that as expected the sensitivity of stellar
evolution properties to the $\czw(\alpha,\gamma)\ose$ rate during the
H- and He-core burning phases of low-mass stars are small. There is a
small dependence of the thermal pulse AGB results on the progenitor
evolution, which however does not change the trends established with
our main set of comparative calculations starting with the same
initial model after the end of He-core burning.

\section{Improving the triple-$\alpha$ rate}
\label{sec:impr}

The recent results of \citet{fynbo:03}, \citep{fynbo:05} have shown
that for $T = 0.1 - 100 \times 10^{8}\kelv$, the triple alpha reaction
rate is dependent essentially on the properties of the $\mem{O^+}$
state at an excitation energy 7.65 MeV in \czw, the Hoyle state. This
reduces the problem of determining the triple alpha rate to
determining the properties of that state.

Because the 7.65 MeV state is a $\mem{O^+}$ state its direct excitation is difficult.
Moreover, the ratio of the radiative width to the total width is
small, $4.13\times 10^{-4}$.  As a result one must determine
$\Gamma_{rad}$ from the relationship
\begin{equation}\label{3alpha2}
\Gamma_{rad}=\Gamma_{\gamma} +
\Gamma_{\pi}=\frac{\Gamma_{\gamma}+
\Gamma_{\pi}}{\Gamma} \frac{\Gamma}{\Gamma_{\pi}} \Gamma_{\pi}
\end{equation}
Here $\Gamma_\gamma$, $\Gamma_\pi$ and $\Gamma$ are the gamma width,
pair width and total width of the 7.65 MeV state.  Each of the three
factors on the right is determined in a separate experiment.  At
present they are known with an accuracy, left to right, of $\pm2.7\%,
\pm9.2\%$, and $\pm6.4\%$.  In all cases, there are several consistent
measurements, so these results can be regarded as robust.

There are two new developments that may significantly improve our
knowledge of $\Gamma_{rad}$, and hence the triple-$\alpha$ reaction
rate, by improving the accuracy of the poorest known quantities: the
pair width $\Gamma_{\pi}$ and the pair branch $\Gamma_{\pi}/\Gamma$ .
The pair width is determined from the transition charge density for
inelastic electron scattering to the 7.6~MeV state. There is a new, as
yet unpublished result \citep{crannell04}, based on a compendium of
extant measurements over a large momentum transfer range, that has a
quoted accuracy of $\pm 2.7\%$. It is difficult to imagine that a more
accurate value of $\Gamma_{\pi}$ can be obtained. On the other hand,
this value is not quite consistent with the earlier values of
$\Gamma_{\pi}$.

The pair branch $\Gamma_{\pi}/\Gamma$ is the least well known
quantity, primarily because it is so small, about $6\times 10^{-6}$. A
new experiment \citep{tur:04}, a Western Michigan University (WMU),
Michigan State University (MSU) collaboration, is underway using the
Tandem accelerator at WMU. The proposed detector is an improved
version of that used by \citet{robertson77}.

In this experiment the 7.6~MeV state in $^{12}$C is excited by
inelastic proton scattering, taking advantage of a strong resonance at
an excitation energy of 10.6 MeV and a scattering angle of 135 degrees
in the lab.  In order to reduce gamma ray backgrounds, a coincidence
is required between a thin plastic- scintillator cylinder surrounding
the target and a large plastic scintillator surrounding both the
target and the cylinder. This arrangement should strongly discriminate
against $\gamma $ ray-backgrounds--gamma rays have only small
probability of interacting in the thin cylinder.  The pair branch is
then given simply by the ratio of the number of positron-electron
pairs detected by plastic scintillator coincidences to the number of
counts in the 7.65~MeV peak in the proton spectrum. An examination of
the systematic uncertainties in the similar Robertson experiment leads
us to estimate that an accuracy of 5\% is achievable.

These two results promise to reduce the uncertainty in the
triple-$\alpha$ rate to about 6\%; as we have seen that will greatly
improve the reliability of predictions of carbon production in AGB
stars.

\section{Discussion}
\label{sec:discussion} 
We have presented a systematic investigation of the propagation of the
rate uncertainties of key nuclear reaction rates into chemical
enrichment predictions of low- and intermediate mass stars that have
reached the thermal pulse AGB phase. We found that the dredge-up in low-mass
stars depends rather sensitively on the adopted reaction rates. The
overall dredge-up of material and, specifically, the yield of
$^{12}$C, has uncertainties of greater than a factor of two owing to the
reaction rate uncertainties. The C/O ratio at the stellar surface has
a similar uncertainty.

Such uncertainties are a problem for many problems of current
astrophysical interest.  The construction of integrated models of
galactic chemical evolution, for example, includes contributions from
stars of all initial masses \citep{timmes:95}, and AGB stars are an
important contributor for some nuclear species.  The enrichment of the
surface abundance with carbon also effects the appearance of AGB stars
in extra-galactic stellar population studies. As the surface abundance
changes from O-dominated ($\textrm{C/O}<1$), to C-rich ($\textrm
{C/O}>1$) the molecular chemistry in the giant's atmosphere changes
considerably \citep{marigo:02}, affecting the star's surface
temperature and thereby its astronomical colors. In older
extra-galactic populations AGB stars are often the brightest stars,
and can probe the population's properties, for example, its age.
Finally, many extremely metal-poor (EMP) stars, which may provide
information on chemical evolution in the early Universe, turn out to
have binary White Dwarf companions \citep{lucatello:04a}. The unusual
abundance patterns of these EMP stars \citep{beers:05} should
correlate with the chemical yields of the White Dwarf progenitors -
the AGB stars at this low metallicity.

Of course nuclear reaction rates are not the only uncertainties in AGB
models. The treatment of convection, and mixing in general, effects
the efficiency of the third dredge-up as well \citep{frost:96,herwig:04c}. Two
separate issues have to be considered. Convection in 1D stellar
evolution models is usually approximated by some variant of the local
mixing-length theory \citep{boehm-vitense:58}.In this ballistic theory
the mean free path of rising and descending blobs has to be
specified.\footnote{Usually one uses the well known stellar parameters
of the sun to calibrate this free parameter, and keeps this value
constant as the evolution progresses. However, multi-D
hydro-simulations have shown that this assumption is not correct
\citep{ludwig:99}, and for evolved giants the mixing-length parameter
may be larger by some substantial fraction.}  Based on the sparse
information on this topic in the literature
\citep{boothroyd:88,ludwig:99,marigo:01b} we roughly estimate that the
mixing-length uncertainty translates into yield uncertainties ranging
from $30\%$ to a factor of a few, depending on initial stellar mass.

Another source of uncertainty of dredge-up predictions is the
treatment of convective overshooting. There is now enough numerical
and experimental proof to claim that convective overshooting takes
place in stellar environments, and that the efficiency of that process
depends on the evolutionary phase \citep{robinson:04}. It appears that
dredge-up predictions are uncertain by a factor of two because of the
poorly known overshooting efficiency.

These two issues related to the modelling of stellar convection have
been viewed as the major source of dredge-up and yield prediction
uncertainties. Our study shows that nuclear reaction rate
uncertainties of two key reactions induce modelling uncertainties of
similar magnitude.  The need to reduce these uncertainties is a
powerful argument for better determinations of the reaction rates of
the $\nvi(\p,\gamma)\ofu$ and triple-$\alpha$ reactions.  Progress in
experimental nuclear physics will have an immediate impact on
astrophysical models that rely on stellar yields.

\acknowledgments{This work was funded in part under the auspices of
the U.S.\ Dept.\ of Energy, and supported by its contract
W-7405-ENG-36 to Los Alamos National Laboratory, the Australian Research Council, 
and by the US NSF
grants PHY01-10253 and PHY02-16783, the latter funding the Joint
Institute for Nuclear Astrophysics (JINA), a NSF Physics Frontier
Center.}


\begin{figure}
\scalebox{0.75}{\includegraphics{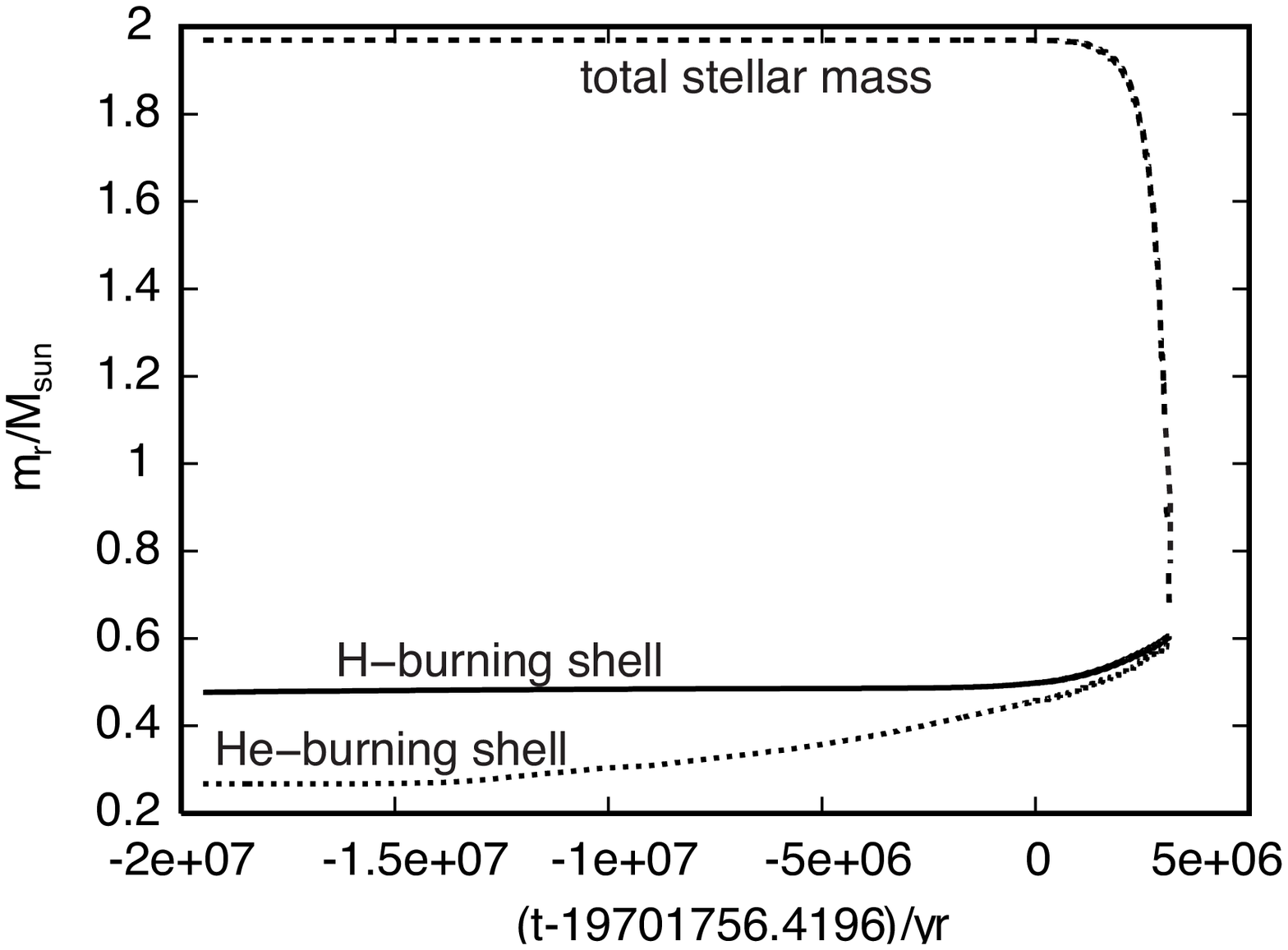}}
\caption{\label{fig:t-M-ET2-long}
Evolution of Lagrange coordinates of stellar surface, H- and
He-burning shell with increasing time. As the star evolves mass loss increases with time  and the total mass decreases.  The lines start at the end of He-core burning. Time has been set to zero at the maximum He-burning  luminosity of the first He-shell flash.}
\end{figure}
\begin{figure}
\includegraphics{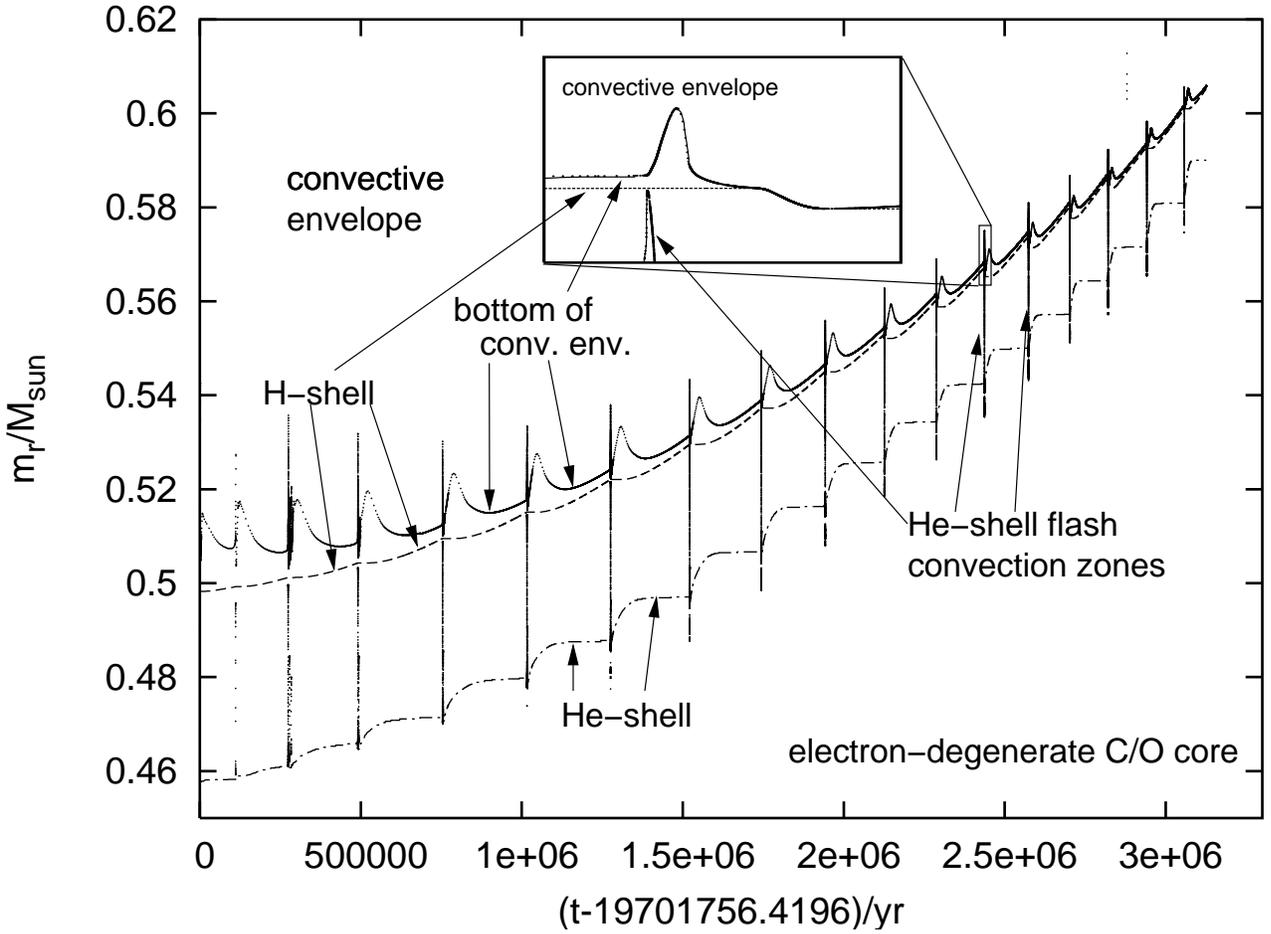}
\caption{\label{fig:t-M-ET2-short} Similar to \abb{fig:t-M-ET2-long},
but only for the AGB phase with He-shell flashes. The inset shows the
detail of an individual dredge-up event. The inset mass-range shown is
from $0.56$ to $0.58\msun$ and the time-range is from
$2.4360\times10^6$ to $2.4375\times10^6\jahre$. More details of AGB
evolution is given in the review article by \citet{herwig:04c}}
\end{figure}

\begin{figure}
\includegraphics{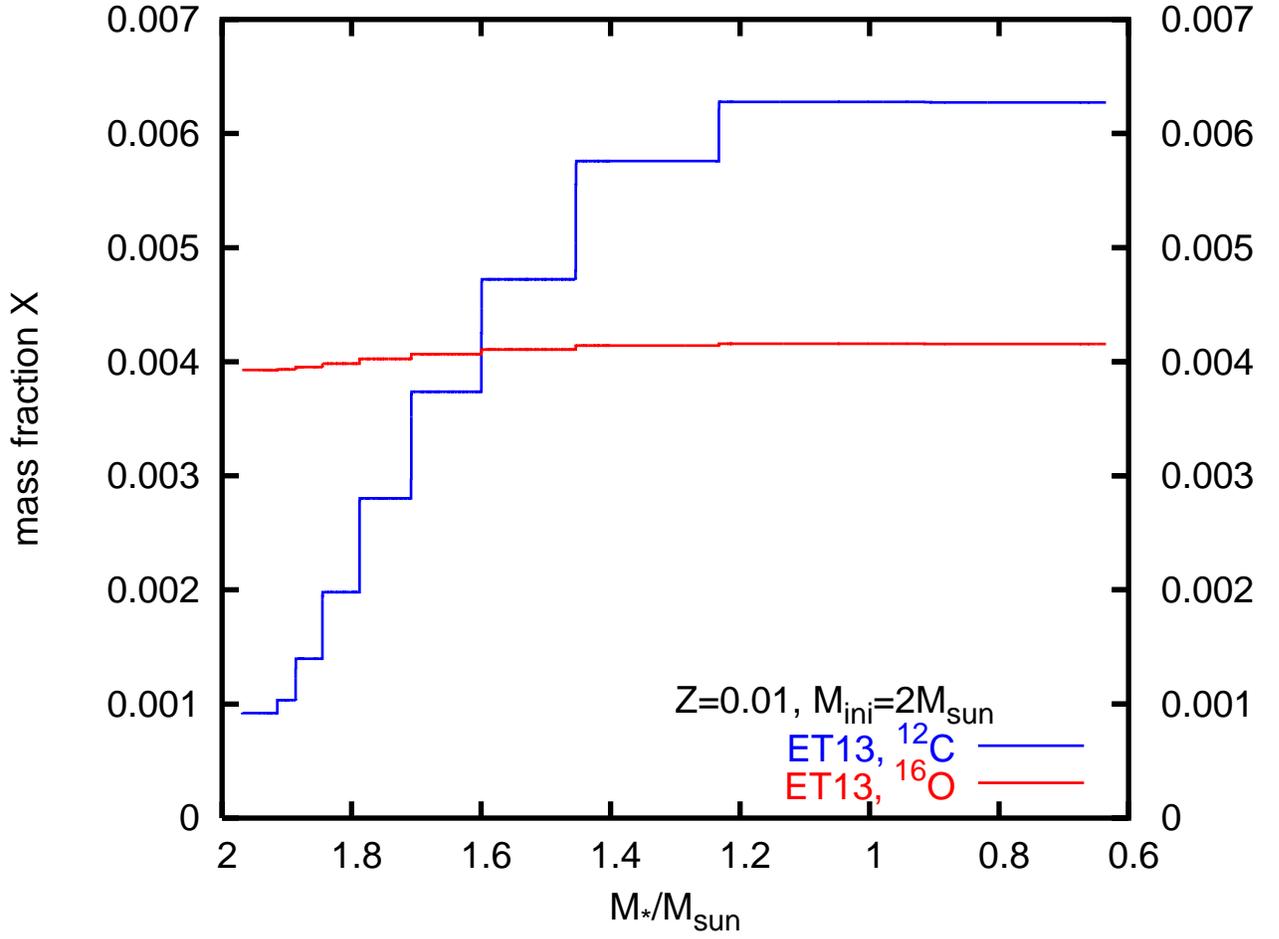}
\caption{\label{fig:t-CO} 
  Evolution of the \czw\ and \ose\ (in mass fraction) in the envelope
  (i.e., at the surface) as a function of total stellar mass (sequence
  ET13). Since the total stellar mass decreases with time the figure
  shows the time evolution of the envelope \czw\ abundance. The
  intershell material that is dredged-up to the envelope contains more
  C than O, which eventually leads to C-star formation
  ($\mem{C/O}>1$). Integration of the surface abundance over the mass
  lost gives the \czw\ yields.  }
\end{figure}

\begin{figure}
\scalebox{0.65}{\includegraphics{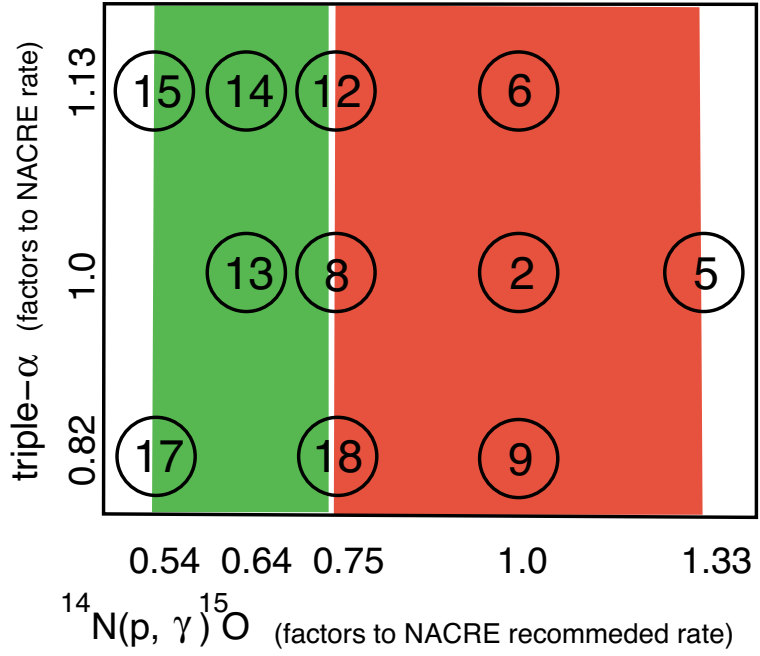}}
\caption{\label{fig:overview}
  Rate selection for stellar evolution He-shell flash calculations.
  The numbers correspond to the sequence numbers in this paper. The
  red (dark) area corresponds to the NACRE range for the $\nvi(\p,\gamma)$
  rate, and the green (light) area covers the revised range for this rate, for
  the temperature range considered in this work (cf.\ 
  \tab{tab:rateselection}). }
\end{figure}

\begin{figure}
\rotatebox{270}{\scalebox{0.5}{\includegraphics{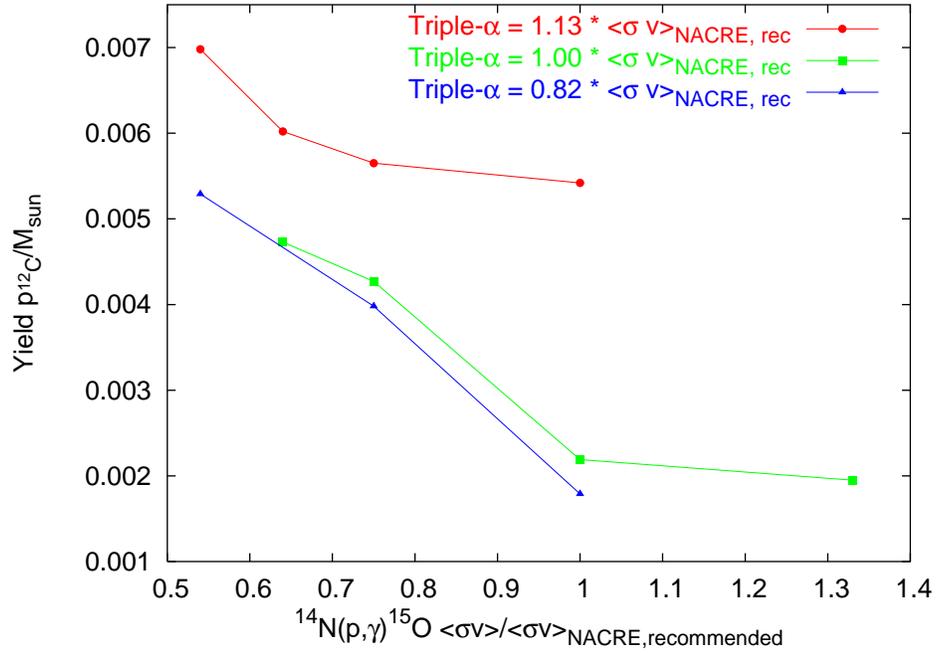}}}
\caption{\label{fig:C12yield} Overview of \czw\ yields as a function
  of the $\nvi(\p,\gamma)$ and triple-$\alpha$ rate, for our
  $M=2\msun$, $Z=0.01$ TP-AGB models as a function of nuclear reaction
  rates. Each point refers to the yield of one full stellar evolution
  model sequence. Lines connect points with the same triple-$\alpha$
  rate. }
\end{figure}

\begin{figure}
\includegraphics{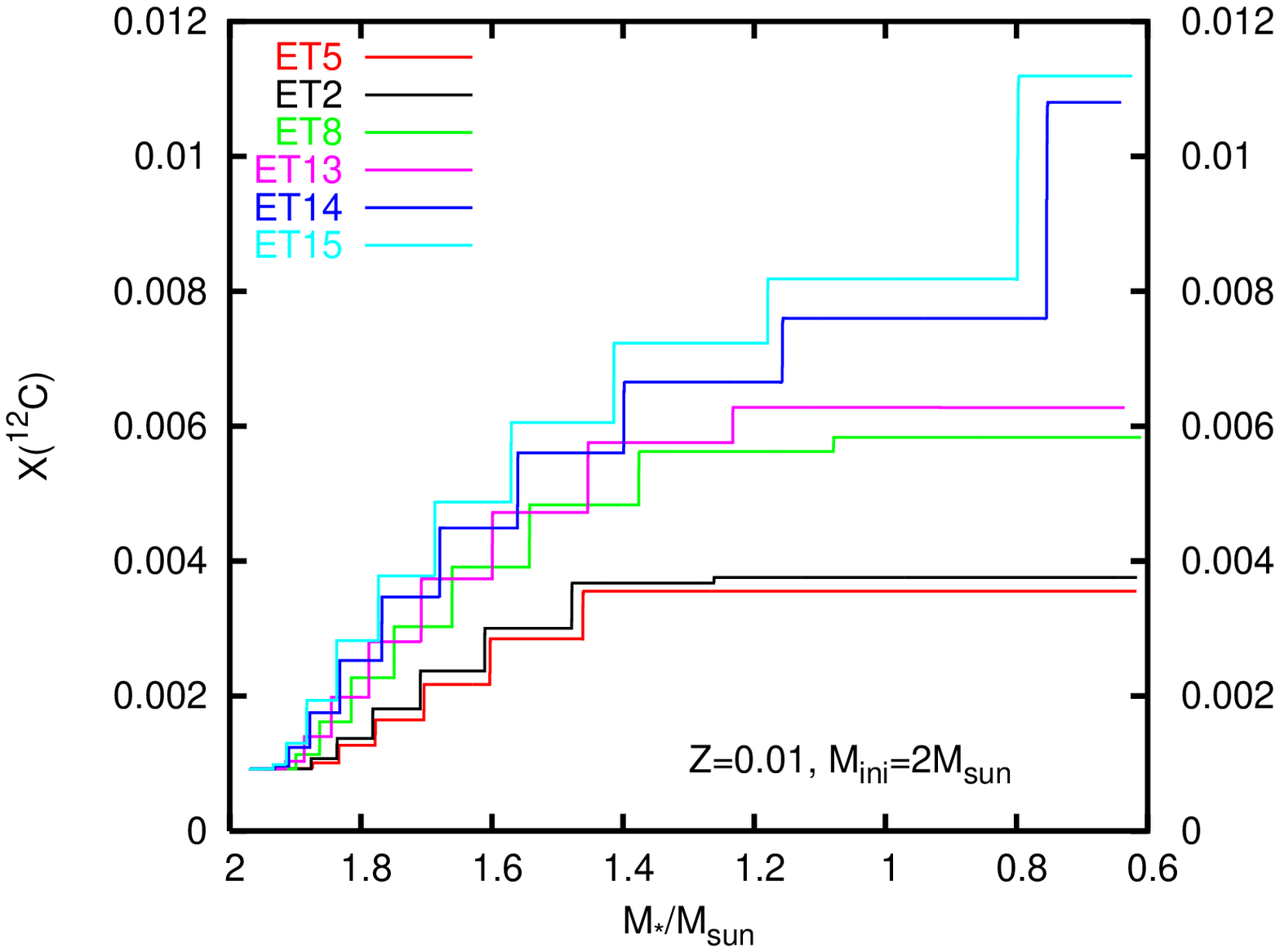}
\caption{\label{fig:C12} 
  Like \abb{fig:t-CO} for sequences with different nuclear reaction
  rates (\abb{fig:overview}). The yields of these sequence are given
  in \tab{tab:sequences1} and \ref{tab:sequences2}.}
\end{figure}

\end{document}